\documentclass{article}
\usepackage{spconf,amsmath,graphicx,hyperref}\hypersetup{hidelinks}
\usepackage{pgfplots}\usepackage{float}\usepackage{placeins}\usepackage{colortbl}\usepackage{multirow}
\pgfplotsset{compat=1.18}
\definecolor{cred}{HTML}{8182BA}
\definecolor{xmk}{HTML}{88C6E2}
\definecolor{mpac}{HTML}{B58581}
\definecolor{cip}{HTML}{A39F9E}
\definecolor{syn}{HTML}{BD4DA3}

\title{SynMark: Semantic-Coordinate Multi-Bit Watermarking during LLM Code Generation}
\name{Rui Xu$^{1}$, Weizhi Liu$^{1}$, Zhaoxia Yin$^{1}$}
\address{$^{1}$East China Normal University, Shanghai, China }
\begin{document}
%\ninept
%
\setlength{\textfloatsep}{2pt plus 1pt minus 1pt}\setlength{\floatsep}{2pt plus 1pt minus 1pt}\setlength{\intextsep}{2pt plus 1pt minus 1pt}\setlength{\abovecaptionskip}{2pt}\setlength{\belowcaptionskip}{0pt}\setlength{\abovedisplayskip}{2pt plus 1pt minus 1pt}\setlength{\belowdisplayskip}{2pt plus 1pt minus 1pt}\setlength{\abovedisplayshortskip}{1pt plus 1pt minus 1pt}\setlength{\belowdisplayshortskip}{2pt plus 1pt minus 1pt}\setlength{\jot}{0.5pt}\maketitle
\begin{abstract}
LLM-generated code is increasingly reused across downstream applications and often undergoes identifier renaming or structural refactoring, making robustness to functionality-preserving edits essential for generative multi-bit code watermarking. Existing methods commonly derive message routes from surface-token prefixes, so such edits can desynchronize routing and degrade watermark recovery. We propose SynMark, which replaces surface-dependent routing with learned semantic coordinates and further calibrates residual recovery errors after transformation. For watermark embedding, Scope learns semantic coordinates from aligned prefixes of functionally equivalent variants for keyed routing and applies a syntax-aware adaptive bias during generation. For watermark extraction, Traced recomputes semantic routes from transformed code and uses reliability-weighted aggregation to suppress unreliable token contributions. Under 100\% identifier renaming, SynMark retains 82.27\% and 83.77\% BitAcc on CodeLlama and DeepSeek-Coder, respectively. Across three code LMs, it also improves clean BitAcc and Pass@1 by up to 4.53 and 11.18 percentage points over the strongest watermarked baselines.
\end{abstract}
\vspace{-0.25em}
\section{Introduction}
\label{sec:intro}

Large language models increasingly generate code that is reused across downstream projects, where identifier renaming, reformatting, and structural refactoring are common~\cite{suresh2024watermarking}. However, external records such as generation logs and platform metadata do not travel with reused code, making provenance difficult to maintain after such edits. Generative watermarking addresses this problem by embedding provenance information directly into token selection, allowing the signal to remain with the generated code when external records are unavailable~\cite{pmlr-v202-kirchenbauer23a,lee-etal-2024-wrote}.

\begin{figure}[t]\centering\vspace{-0.1em}\includegraphics[width=1\linewidth]{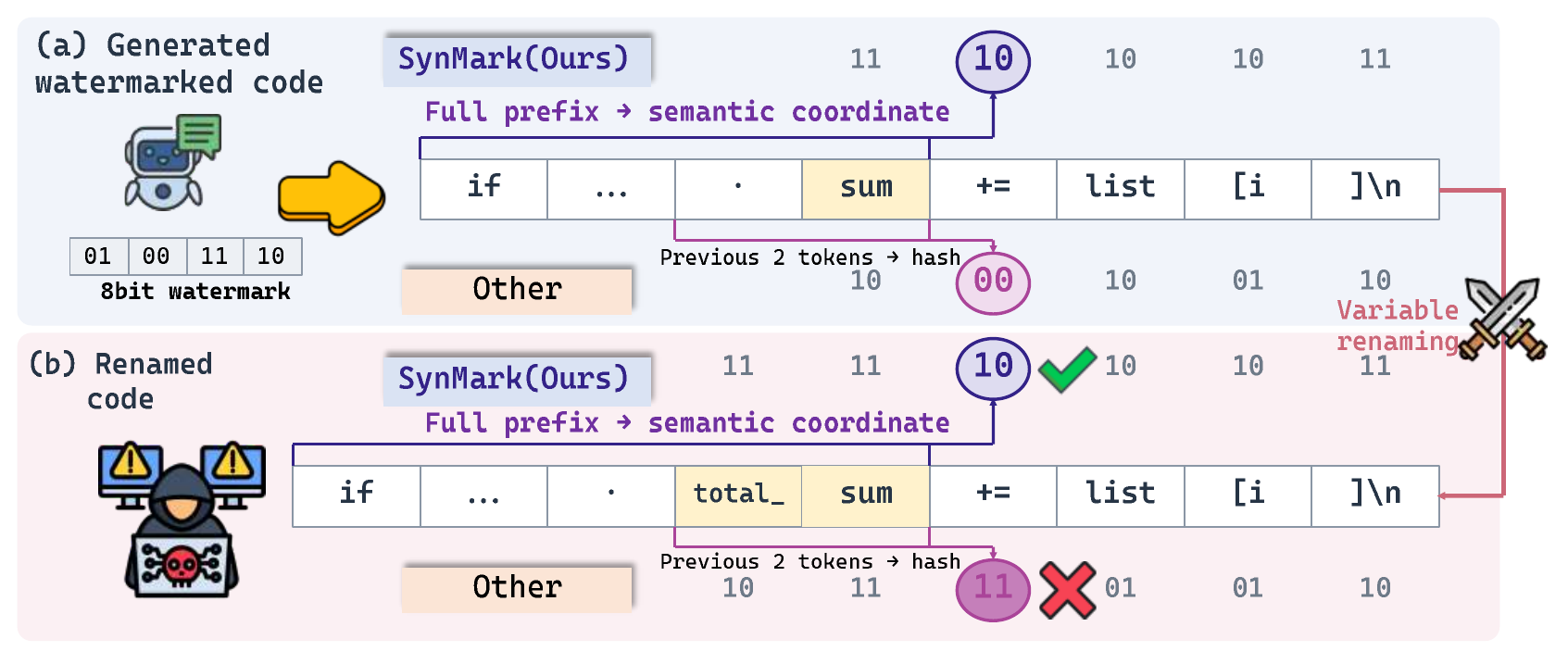}\vspace{-0.2em}\caption{Token-based and semantic-coordinate routing under identifier renaming. Binary strings are two-bit message symbols; token-based routing uses nearby tokens, while SynMark routes from a semantic coordinate of the full prefix.}\label{fig:intro_overview}\vspace{-0.1em}\end{figure}

However, existing generative watermarking methods struggle to trace code fragments after semantics-preserving transformations. They commonly derive keyed message routes and candidate sets directly from surface-token prefixes~\cite{yoo-etal-2024-advancing,jiang2026credid,xu-etal-2026-xmark}. Transformations can then impair recovery at two coupled levels: \textit{(i) modified prefixes may cause the detector to reconstruct routes or candidate sets different from those used during generation, as illustrated in Fig.~\ref{fig:intro_overview}}; and \textit{(ii) even under a correct route, renamed, inserted, or rewritten tokens may produce unreliable recovery signals}. The former changes message-block assignment, while the latter degrades token-level recovery after routing. Increasing the watermark bias cannot correct either failure and may alter syntax-sensitive token choices, reducing functional correctness. Robust recovery therefore requires both stable semantic routing references and calibration of the residual transformation-induced variation in token-level recovery signals.

\par
\par We propose SynMark, a generative multi-bit watermarking framework that embeds watermark messages during LLM code generation. Unlike existing methods that directly couple keyed routing to surface-token prefixes, Scope first maps each prefix to a \textbf{discrete semantic coordinate} learned from aligned, functionally equivalent code variants and performs keyed block assignment on this intermediate reference. During generation, Scope also applies a \textbf{syntax-aware adaptive bias} to limit watermark-induced changes at syntax-sensitive positions. Since semantic routing reduces but does not eliminate transformation-induced variation, Traced recomputes semantic routes during extraction and calibrates the remaining token-level recovery signals through \textbf{candidate-score correction and reliability-weighted aggregation}.
 \par We evaluate SynMark on HumanEval with OpenCoder, CodeLlama, and DeepSeek-Coder against MPAC, CredID, XMark, and CodeIP. SynMark improves 8-bit message recovery across all three generators, retains higher Pass@1 than competing watermarked methods, and achieves stronger recovery under identifier renaming and random code transformations.\par Our main contributions are summarized as follows:
\begin{itemize}
    \item \textbf{Semantic-coordinate routing.} We introduce a causal routing mechanism that learns discrete semantic coordinates from aligned prefixes of functionally equivalent code, reducing dependence on surface-token references during both embedding and extraction.

    \item \textbf{Syntax-aware watermark embedding.} We develop an adaptive logit bias that limits watermark-induced changes at syntax-sensitive positions while retaining the default embedding strength elsewhere.

    \item \textbf{Reliability-weighted extraction.} We design Traced to compensate for residual transformation-induced drift in token-level recovery signals through candidate-score correction and reliability-weighted aggregation.
\end{itemize}

\vspace{-0.5em}
\section{Related Work}
\label{sec:related_work}
Generative text watermarking partitions vocabularies and biases token probabilities with a secret key~\cite{pmlr-v202-kirchenbauer23a}. Recent methods improve robustness via semantic representations~\cite{ren-etal-2024-robust} or uncertainty-adaptive strength~\cite{pmlr-v235-liu24e}. They mainly target natural language, leaving code syntax constraints and multi-bit route synchronization underexplored.

Multi-bit watermarking links generation positions to message symbols through keyed vocabulary partitions. MPAC routes symbols from preceding tokens~\cite{yoo-etal-2024-advancing}, CredID combines prefix cues with model logits~\cite{jiang2026credid}, and XMark uses keyed vocabulary permutations for reliable message recovery~\cite{xu-etal-2026-xmark}. For code generation, prior work studies watermark-aware token constraints~\cite{lee-etal-2024-wrote}, grammar-guided multi-bit encoding~\cite{guan-etal-2024-codeip}, syntax-critical decisions~\cite{kim-etal-2026-marking}, and user-level tracing~\cite{zhang-etal-2025-clmtracing}. These methods improve detection and tracing, yet token-level routing, grammar constraints, and fixed candidate construction remain sensitive to renaming, reformatting, and structural rewrites. SynMark uses semantic-coordinate routing and reliability-weighted extraction to improve recovery after semantics-preserving transformations.

\begin{figure}[t]\centering\vspace{-0.4em}\includegraphics[width=1\linewidth]{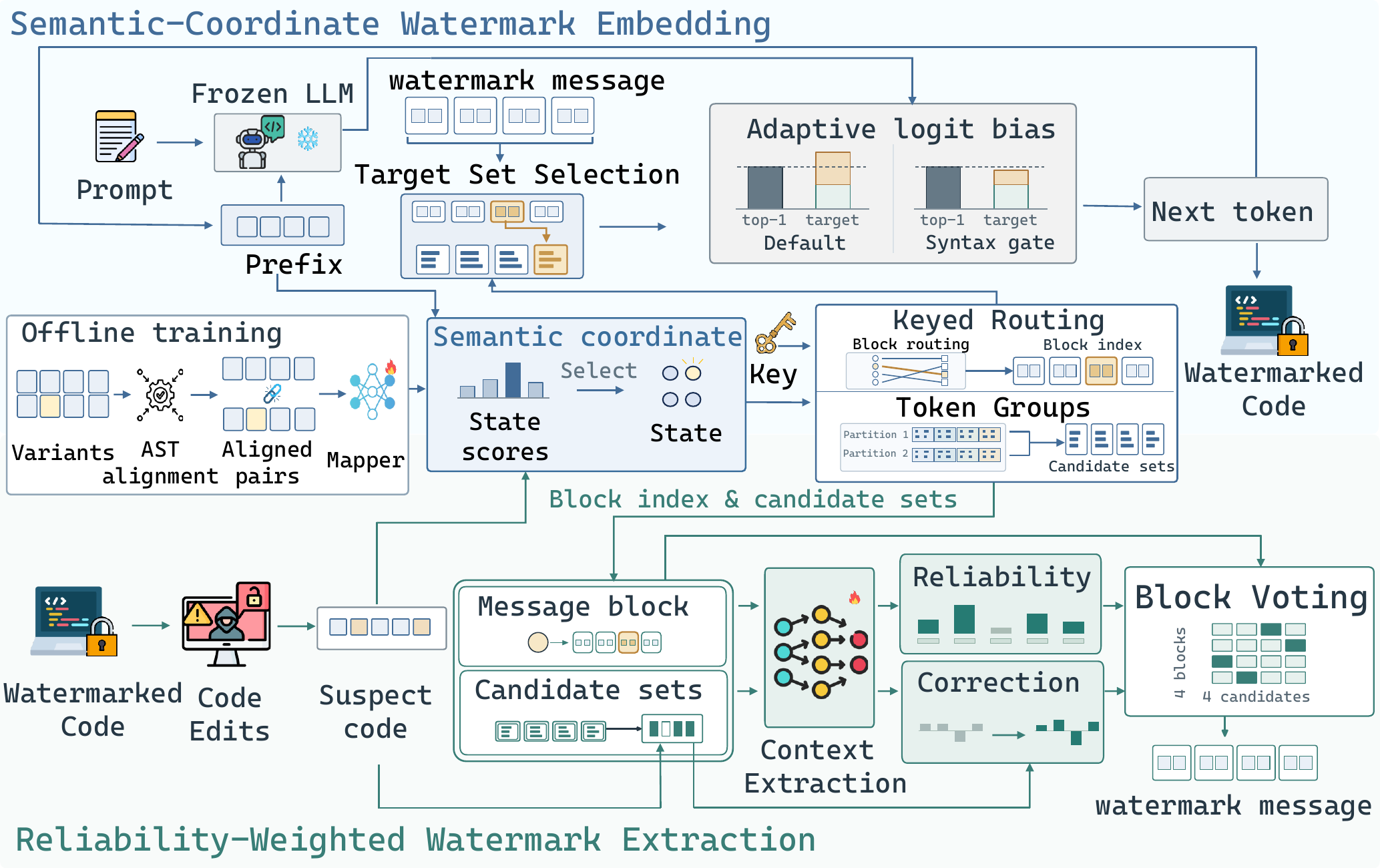}\vspace{-0.3em}\caption{Overview of SynMark. Scope embeds during generation; Traced recomputes routes for reliability-weighted extraction from transformed code.}\label{fig:framework}\end{figure}
\section{Proposed Method}
\label{sec:method}

\vspace{-0.25em}\subsection{Problem Formulation and Threat Model}
Given a prompt $x$, an $L$-bit watermark message $m\in\{0,1\}^L$, and a secret key $\kappa$, SynMark generates $y=\mathcal G_\kappa(x,m)$ and recovers $\hat m=\mathcal D_\kappa(\widetilde y)$ after edits $\widetilde y=\mathcal T(y)$. For code expected to carry a watermark, the detector outputs $\hat m$, and we evaluate bit-level recovery. For an even payload length $L$, we split $m$ into $B=L/2$ two-bit blocks, where $q_{b,\mathrm{hi}},q_{b,\mathrm{lo}}\in\{0,1\}$ and $q_b=2q_{b,\mathrm{hi}}+q_{b,\mathrm{lo}}\in[Q]$ with $Q=4$. The mapper uses $S=B$ states, and the keyed bijection $\pi_\kappa:[S]\to[B]$ assigns states to message blocks. Figure~\ref{fig:framework} summarizes Scope embedding and Traced extraction; Traced reuses the tokenizer, frozen embeddings, and trained mapper for correction and reliability weighting.

\textbf{Threat model.} We consider post-generation attacks where an adversary edits watermarked code $y$ to weaken the watermark while preserving functionality, without access to the secret key $\kappa$ or detector feedback. These edits include identifier renaming, semantics-preserving structural refactoring, and comment/dead-code insertion. The detector recovers the watermark message from $\widetilde y$. Full re-synthesis, code regeneration, functionality-changing edits, and key compromise are out of scope.

\vspace{-0.4em}\subsection{Semantic-Coordinate Watermark Embedding}
\label{subsec:scope}
Scope trains a causal mapper $f_\phi$ to assign aligned prefixes of functionality-preserving variants to shared discrete coordinates. For each snippet $a$, we construct $\widetilde a$. During transformation, we retain correspondences between preserved AST nodes and map their end offsets to BPE boundaries using tokenizer offsets, yielding $\mathcal A=\{(t,t')\}$; inserted or deleted nodes without counterparts are omitted. Mini-batch Sinkhorn converts clean-prefix logits into balanced soft assignments $\hat{\mathbf p}_{t,t'}$, shared as stop-gradient targets with aligned transformed prefixes. Let $w_{t,t'}=\omega_{\mathrm{tr}}$ if $a_t\neq\widetilde a_{t'}$, and $w_{t,t'}=1$ otherwise. The mapper uses frozen generator embeddings, normalization/projection layers, a GRU prefix encoder, and an $S$-way classifier. Its pair-weighted loss is:
\begin{equation}
\begin{aligned}
\mathcal L_{\mathrm{pair}}
&=\frac{1}{\sum_{(t,t')\in\mathcal A}w_{t,t'}}
\sum_{(t,t')\in\mathcal A}w_{t,t'}\Big[
\operatorname{CE}\!\left(f_\phi(a_{<t}),\hat{\mathbf p}_{t,t'}\right)\\
&\qquad+
\operatorname{CE}\!\left(f_\phi(\widetilde a_{<t'}),\hat{\mathbf p}_{t,t'}\right)
\Big].
\end{aligned}
\end{equation}
The Scope objective is
$\mathcal L_{\mathrm{scope}}
=\mathcal L_{\mathrm{pair}}
+\mathcal L_{\mathrm{repr}}
+\mathcal L_{\mathrm{use}}
+0.25\mathcal L_{\mathrm{seq}}
+0.1\mathcal L_{\mathrm{iso}}
+0.05\mathcal L_{\mathrm{conf}}$.
Here, $\mathcal L_{\mathrm{repr}}$ aligns paired representations,
$\mathcal L_{\mathrm{use}}$ and $\mathcal L_{\mathrm{seq}}$ balance
state usage globally and within sequences, and
$\mathcal L_{\mathrm{iso}}$ and $\mathcal L_{\mathrm{conf}}$
regularize representation geometry and prediction confidence.
AST alignment is used only during training.

Let $\mathcal V_*$ be the non-special-token vocabulary. At position $t$, $z_t=\arg\max_{s\in[S]}[f_\phi(y_{<t})]_s$ and $b_t=\pi_\kappa(z_t)$. For $j\in\{1,2\}$, $h_{\kappa,j}:[S]\times\mathcal V_*\to[Q]$ defines $P_{\kappa,j}(z,q)=\{v\in\mathcal V_*:h_{\kappa,j}(z,v)=q\}$; the same $\kappa$ parameterizes $\pi_\kappa,h_{\kappa,1},h_{\kappa,2}$. SynMark sets $G_\kappa(z,q)=\mathcal V_*\setminus\bigcup_{j=1}^{2}P_{\kappa,j}(z,q)$ and $\mathcal{W}_t=G_\kappa(z_t,q_{b_t})$. Complementary coding yields multi-hot candidate indicators: a token supports $q$ only outside the $q$-th shard in both partitions~\cite{xu-etal-2026-xmark}. Detection recomputes coordinates and candidate sets from transformed-code prefixes, so block assignment does not use absolute token positions. This anchors block assignment to learned coordinates rather than surface tokens, reducing its direct dependence on token identity. Traced then handles the remaining variation in token-level recovery signals.

Let $\ell_t(v)$ be the frozen LLM pre-watermark logit, $\mathcal C_{\mathrm{syn}}$ the syntax-sensitive tokens (keywords, operators, whitespace/formatting), and $v_t^*=\arg\max_v\ell_t(v)$ the original top-1 token. Define the target margin as
\begin{equation}
d_t=\ell_t(v_t^*)-\max_{v\in \mathcal{W}_t}\ell_t(v).
\end{equation}
Let $\delta>\tau>0$ denote the default bias and margin floor. At position $t$, $\epsilon_t=\ell_t(v_t^*)-\operatorname{nextafter}\!\left(\ell_t(v_t^*),-\infty\right)$ is the dtype-dependent numerical slack; the adaptive bias is:
\begin{equation}
\begin{array}{@{}l@{\;}c@{\;}l@{}}
g_t & = & \mathbf 1\left[v_t^*\in\mathcal C_{\mathrm{syn}}\land v_t^*\notin \mathcal{W}_t\land \tau\leq d_t<\delta\right],\\
\Delta_t & = & g_t\max\{0,\min(\delta,d_t-\epsilon_t)\}+(1-g_t)\delta,\\
\ell'_t(v) & = & \ell_t(v)+\Delta_t\mathbf{1}[v\in \mathcal{W}_t].
\end{array}
\end{equation}
For active gates, the clipped bias prevents any boosted target token from overtaking the original top-1 token. When $d_t<\tau$, the strongest target token is already close to the original top-1, so the default bias is retained to preserve watermark strength.

\vspace{-0.4em}\subsection{Reliability-Weighted Watermark Extraction}
\label{subsec:decoder}
At detection, Traced recomputes $(z_t,b_t)$ from $\widetilde y_{<t}$ using $f_\phi$ and $\pi_\kappa$, rebuilds $G_\kappa(z_t,q)$, and forms $o_{t,q}=\mathbf 1[\widetilde y_t\in G_\kappa(z_t,q)]$. It runs only during extraction and does not affect generation. We define
\begin{equation}
\begin{aligned}
o_t&=(o_{t,0},\ldots,o_{t,Q-1}),\\
\bar o_t&=\frac{1}{Q}\sum_{q'=0}^{Q-1}o_{t,q'}.
\end{aligned}
\end{equation}
For transformed-code length $T$, $\psi_t$ concatenates learned embeddings of the token hash, sanitized token, lexical type, and route $(z_t,b_t)$ with $\rho_t=t/T$ and $o_t$. Sanitization maps identifiers and literals to shared categories. Surface and position features contextualize only Traced; $b_t=\pi_\kappa(z_t)$ determines block assignment. A BiGRU produces $h_t=\operatorname{BiGRU}(\psi_{1:T})_t$, and a shared network $F_c$ outputs a scalar correction for each candidate $q$:
\begin{equation}
\begin{aligned}
c_{t,q} &= \tanh F_c\!\left([h_t;u_q;o_{t,q};o_{t,q}-\bar o_t]\right),\\
\widetilde c_{t,q} &= c_{t,q}-\frac{1}{Q}\sum_{q'=0}^{Q-1}c_{t,q'},\\
e_{t,q} &= o_{t,q}-\bar o_t+\widetilde c_{t,q}.
\end{aligned}
\end{equation}
Here $u_q$ is the learned embedding of candidate symbol $q$; $F_c$ combines context, raw membership, and centered membership to adjust candidate-relative scores.

\begin{table}[t]
\caption{Comparison of watermark recovery and functional correctness on HumanEval across three code LMs, reported as BitAcc/Pass@1 (\%).}
\label{tab:main}
\centering
\small\setlength{\tabcolsep}{2.0pt}
\renewcommand{\arraystretch}{1.08}
\begin{tabular}{@{}lccc@{}}
\hline
Method & OpenCoder & CodeLlama & DeepSeek \\
\hline
Unwatermarked & --/40.24 & --/21.95 & --/34.76 \\
CredID~\cite{jiang2026credid} & 68.60/26.83 & 81.86/7.32 & 81.10/16.46 \\
XMark~\cite{xu-etal-2026-xmark} & 75.00/25.61 & 83.99/8.54 & 79.19/6.71 \\
MPAC~\cite{yoo-etal-2024-advancing} & 77.74/23.78 & 77.29/2.44 & 82.93/12.20 \\
CodeIP~\cite{guan-etal-2024-codeip} & 77.31/26.83 & 69.33/6.10 & 69.56/2.44 \\
\hline
\rowcolor{gray!15}\textbf{SynMark (Ours)} & \textbf{80.26/36.99} & \textbf{88.52/19.72} & \textbf{86.99/27.03} \\
\hline
\end{tabular}
\end{table}

The reliability branch sets $r_t=2\sigma(\mathbf w_r^\top\operatorname{LN}(h_t)+\beta_r)\in(0,2)$, with $\mathbf w_r$ matching $h_t$ and scalar $\beta_r$; initialization $\mathbf w_r=\mathbf 0,\beta_r=0$ gives $r_t=1$ before reliability training. For block $b$, Traced computes
\begin{equation}
\Sigma_{b,q}=
\frac{\sum_{t:b_t=b}r_t e_{t,q}}
{\sqrt{\max\!\left(1,\sum_{t:b_t=b}r_t^2\right)}}.
\end{equation}
The denominator normalizes reliability weights with a unit lower bound on their squared sum; empty blocks use zero scores. We then compute $p_{b,q}=\operatorname{softmax}_q(\Sigma_{b,q})$ and $\hat q_b=\arg\max_q p_{b,q}$.

During Traced training, the fixed Scope module embeds sampled messages, and clean/transformed inputs share labels. With $r_t=1$, the correction stage trains the feature and candidate embeddings, BiGRU, and $F_c$ using $\mathcal L_{\mathrm{corr}}=\mathcal L_{\mathrm{sym}}\allowbreak+\lambda_{\mathrm{bit}}\mathcal L_{\mathrm{bit}}\allowbreak+\lambda_{\mathrm{res}}\mathcal L_{\mathrm{res}}$. The reliability stage freezes these modules and trains only the reliability branch initialized at $r_t=1$ using $\mathcal L_{\mathrm{rel}}=\mathcal L_{\mathrm{sym}}\allowbreak+\lambda_{\mathrm{bit}}\mathcal L_{\mathrm{bit}}$. For $q_b=2q_{b,\mathrm{hi}}+q_{b,\mathrm{lo}}$, define $P_b^{\mathrm{hi}}(1)=p_{b,2}+p_{b,3}$ and $P_b^{\mathrm{lo}}(1)=p_{b,1}+p_{b,3}$. We use $\mathcal L_{\mathrm{sym}}=-B^{-1}\allowbreak\sum_{b=0}^{B-1}\log p_{b,q_b}$, $\mathcal L_{\mathrm{bit}}=(2B)^{-1}\allowbreak\sum_b\allowbreak\sum_{\eta\in\{\mathrm{hi},\mathrm{lo}\}}\allowbreak\operatorname{BCE}(P_b^{\eta}(1),q_{b,\eta})$, and $\mathcal L_{\mathrm{res}}=(TQ)^{-1}\allowbreak\sum_{t=1}^{T}\allowbreak\sum_{q=0}^{Q-1}\widetilde c_{t,q}^{\,2}$. The residual penalty prevents corrections from dominating raw membership signals. Inference converts each selected symbol into two bits.

\begin{figure}[t]\centering\resizebox{\columnwidth}{!}{\begin{tikzpicture}\begin{axis}[name=codellama,width=4.25cm,height=3.45cm,scale only axis,title={(a) CodeLlama-7B},title style={font=\small,yshift=-0.4ex},ylabel={BitAcc (\%)},xmin=22,xmax=103,ymin=50,ymax=90,xtick={25,50,75,100},ytick={50,60,70,80,90},ymajorgrids=true,grid style={gray!20},tick label style={font=\small},label style={font=\small},line width=1pt,mark size=2pt,]\addplot[color=cred,dashed,mark=square*] coordinates {(25,77.27)(50,77.05)(75,68.64)(100,65.23)};\addplot[color=xmk,dash dot,mark=triangle*] coordinates {(25,80.62)(50,78.75)(75,75.62)(100,70.83)};\addplot[color=mpac,dotted,mark=diamond*] coordinates {(25,78.72)(50,79.05)(75,78.38)(100,76.35)};\addplot[color=cip,dashed,mark=triangle*] coordinates {(25,57.06)(50,53.63)(75,57.66)(100,53.83)};\addplot[color=syn,very thick,mark=*] coordinates {(25,87.50)(50,87.25)(75,85.76)(100,82.27)};\end{axis}\begin{axis}[name=deepseek,at={(codellama.east)},anchor=west,xshift=0.36cm,width=4.25cm,height=3.45cm,scale only axis,title={(b) DeepSeek-Coder-6.7B},title style={font=\small,yshift=-0.4ex},xmin=22,xmax=103,ymin=50,ymax=90,xtick={25,50,75,100},ytick={50,60,70,80,90},yticklabels={},ymajorgrids=true,grid style={gray!20},tick label style={font=\small},label style={font=\small},line width=1pt,mark size=2pt,]\addplot[color=cred,dashed,mark=square*] coordinates {(25,65.52)(50,64.72)(75,62.70)(100,58.06)};\addplot[color=xmk,dash dot,mark=triangle*] coordinates {(25,73.04)(50,71.08)(75,69.85)(100,63.73)};\addplot[color=mpac,dotted,mark=diamond*] coordinates {(25,77.50)(50,76.29)(75,74.24)(100,70.38)};\addplot[color=cip,dashed,mark=triangle*] coordinates {(25,51.39)(50,51.85)(75,51.39)(100,51.39)};\addplot[color=syn,very thick,mark=*] coordinates {(25,86.18)(50,85.96)(75,83.77)(100,83.77)};\end{axis}\end{tikzpicture}}\par\vspace{-0.15em}\centerline{\small Renamed identifiers (\%)}\par\vspace{0.05em}\centerline{\small\tikz[baseline=-0.5ex]{\draw[cred,dashed,line width=0.9pt] (0,0)--(0.24,0);}\,CredID\quad\tikz[baseline=-0.5ex]{\draw[xmk,dash dot,line width=0.9pt] (0,0)--(0.24,0);}\,XMark\quad\tikz[baseline=-0.5ex]{\draw[mpac,dotted,line width=0.9pt] (0,0)--(0.24,0);}\,MPAC\quad\tikz[baseline=-0.5ex]{\draw[cip,dashed,line width=0.9pt] (0,0)--(0.24,0);}\,CodeIP\quad\tikz[baseline=-0.5ex]{\draw[syn,very thick,line width=1.1pt] (0,0)--(0.24,0);}\,SynMark}\vspace{-0.2em}\caption{BitAcc under identifier renaming on two code LMs.}\label{fig:var_robustness}
\end{figure}
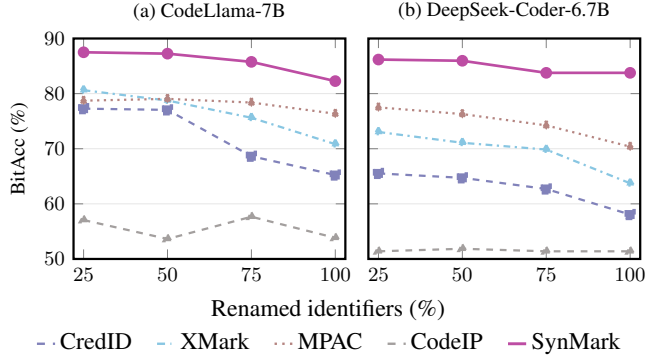

\section{Experiments}
\label{sec:experiments}
\subsection{Experimental Setup}
\label{subsec:setup}
We train generator-specific Scope and Traced modules on MBPP (320/80 train/validation)~\cite{austin2021programsynthesislargelanguage} and evaluate on HumanEval~\cite{chen2021evaluatinglargelanguagemodels} with OpenCoder-1.5B~\cite{huang-etal-2025-opencoder}, CodeLlama-7B~\cite{roziere2023codellama}, and DeepSeek-Coder-6.7B~\cite{guo2024deepseekcoder}. We compare with CredID~\cite{jiang2026credid}, XMark~\cite{xu-etal-2026-xmark}, MPAC~\cite{yoo-etal-2024-advancing}, and CodeIP~\cite{guan-etal-2024-codeip} under the same 8-bit payload and generation configuration, using their recommended watermark strengths. Results are averaged over three seeds. Robustness is evaluated on successfully transformed, functionality-preserving HumanEval outputs using the same transformation families as those used in Scope training: identifier renaming and 1--3 sequential rewrites of conditionals, loops, updates, and declarations/initialization, together with dead-code and comment insertion. We set $(\delta,\tau,\omega_{\mathrm{tr}})=(4,2,4)$ and $(\lambda_{\mathrm{bit}},\lambda_{\mathrm{res}})=(0.25,0.01)$.

\subsection{Watermark Accuracy and Fidelity}
\label{subsec:accuracy}
Table~\ref{tab:main} compares clean watermark recovery and
code fidelity. SynMark achieves the highest BitAcc and Pass@1 among the evaluated watermarked methods across all three generators. The largest gains occur on CodeLlama, where SynMark improves BitAcc and Pass@1 by 4.53 and 11.18 percentage points, respectively, over XMark, the strongest watermarked baseline in both metrics. SynMark also narrows the fidelity gap to unwatermarked generation, reaching
19.72\% Pass@1 compared with 21.95\% on CodeLlama.
These results show that improved bit recovery need not
come at a greater cost to functional correctness.
This is consistent with SynMark's adaptive embedding
design, which limits watermark intervention at gated
syntax-sensitive positions by preserving the original
top-1 token ranking.

\subsection{Robustness under Code Transformations}
Figure~\ref{fig:var_robustness} compares BitAcc under identifier renaming on CodeLlama and DeepSeek-Coder. SynMark consistently outperforms all baselines across renaming ratios, retaining 82.27\% and 83.77\% BitAcc under 100\% renaming. These results show that SynMark is less sensitive to identifier changes. Semantic-coordinate routing reduces dependence on token identities, while Traced calibrates the remaining token-level recovery signals.

Table~\ref{tab:robustness} evaluates broader semantics-preserving edits on OpenCoder. SynMark achieves the highest BitAcc in every attack group, improving the averages under identifier renaming and code transformation attacks by 8.62\% and 5.85\% over the strongest corresponding baselines. It also performs best under dead-code and comment insertion. These gains reflect the complementary roles of Scope and Traced: Scope reduces routing dependence on surface-token identities, while Traced corrects and downweights unreliable recovery signals introduced by structural rewrites and token insertions.

\begin{table}[t]
\caption{Robustness comparison on OpenCoder under identifier renaming, random code transformations, dead-code insertion, and comment insertion, reported as BitAcc (\%).}
\label{tab:robustness}
\centering
\setlength{\tabcolsep}{2.0pt}
\small
\renewcommand{\arraystretch}{1.08}\begin{tabular}{@{}lcccc@{}}
\hline
Method & Var. & Trans. & Dead & Comment \\
\hline
CredID~\cite{jiang2026credid} & 58.35 & 51.65 & 68.14 & 69.02 \\
XMark~\cite{xu-etal-2026-xmark} & 67.51 & 55.09 & 72.16 & 74.92 \\
MPAC~\cite{yoo-etal-2024-advancing} & 71.45 & 60.66 & 76.98 & 76.37 \\
CodeIP~\cite{guan-etal-2024-codeip} & 61.22 & 52.34 & 65.03 & 64.13 \\
\rowcolor{gray!15}\textbf{SynMark (Ours)} & \textbf{77.61} & \textbf{64.21} & \textbf{79.71} & \textbf{79.62} \\
\hline
\end{tabular}
\end{table}

\subsection{Ablation Study}
\label{subsec:ablation}
For the ablation, removing Scope reverts to XMark's surface-token routing with a fixed watermark bias, while removing Traced uses standard block voting~\cite{xu-etal-2026-xmark}. Table~\ref{tab:ablation} shows that adding Scope improves Pass@1 by 11.38 percentage points and Trans. Avg. BitAcc by 5.16 points, while Clean and Var. Avg. remain comparable. These results are consistent with Scope's two generation-side mechanisms: semantic-coordinate routing reduces dependence on surface-token references, while the syntax-aware bias limits watermark interference with code generation. With Scope fixed, Traced further improves Clean, Var., and Trans. Avg. BitAcc by 5.87, 10.10, and 3.96 points, respectively, without affecting Pass@1, indicating that extraction-side calibration remains useful for residual recovery errors. We omit the configuration without Scope but with Traced because Traced relies on Scope-derived routes. Overall, the ablation supports the complementary roles of Scope embedding and Traced extraction in improving fidelity and recovery under transformations.

\begin{table}[!ht]
\caption{Ablation results on OpenCoder (\%). ``w/'' and ``w/o'' denote with and without each module, respectively.}
\label{tab:ablation}
\centering
\small\setlength{\tabcolsep}{2.0pt}
\renewcommand{\arraystretch}{1.08}\begin{tabular}{@{}llcccc@{}}
\hline
Scope & Traced & Clean & Var Avg. & Trans. Avg. & Pass@1 \\
\hline
\rowcolor{gray!15}\textbf{w/} & \textbf{w/} & \textbf{80.26} & \textbf{77.61} & \textbf{64.21} & \textbf{36.99} \\
w/ & w/o & 74.39 & 67.51 & 60.25 & 36.99 \\
w/o & w/o & 75.00 & 67.51 & 55.09 & 25.61 \\
\hline
\end{tabular}
\end{table}

\section{Conclusion}SynMark embeds multi-bit watermark messages during LLM code generation with semantic-coordinate routing, syntax-aware adaptive bias, and reliability-weighted extraction. On HumanEval, SynMark improves bit recovery, functional correctness, and robustness over evaluated watermarked baselines across three code LMs.
\begingroup\small
\section{Acknowledgment}
This work was supported by the National Natural Science Foundation of China (Grant No. 62472177). The authors declare no conflicts of interest.

\section{Compliance with Ethical Standards}
This study involved no human participants or animals and required no ethical approval.

\bibliographystyle{IEEEbib}
\bibliography{refs}

\endgroup\end{document}